\documentclass[final,1p,times]{elsarticle}

\usepackage{amsmath,amssymb,bm}
\usepackage{comment}
\begin{document}

%%%%%%%%%%%%%%%%%%%%%%%%%%%%%%%
%%%%%%%%%%%%%%%%%%%%%%%%%%%%%%%
\begin{flushright}
KEK-TH-2171 and J-PARK-TH-0206
\end{flushright}
%%%%%%%%%%%%%%%%%%%%%%%%%%%%%%%
%%%%%%%%%%%%%%%%%%%%%%%%%%%%%%%

\begin{frontmatter}



\title{The physics of helical electron beam in a uniform magnetic field
as a testing ground of gauge principle}

%\title{Does the gauge principle permit the observation of the canonical orbital
%angular momentum of the electron in the Landau problem ?}

%% \adress[label1]{}

\author[label1,label2]{Masashi Wakamatsu\corref{cor1}}
\ead{wakamatu@post.kek.jp}
\author[label1]{Yoshio~Kitadono}
%\ead{kitadono@impcas.ac.cn}
\author[label1]{Liping Zou}
\author[label3]{Pengming Zhang}
%\ead{zhpm@impcas.ac.cn}

\cortext[cor1]{corresponding author}

\address[label1]{Institute of Modern Physics, Chinese Academy of Sciences,\\
Lanzhou, 730000, People's Republic of China}
\address[label2]{KEK Theory Center, Institute of Particle and Nuclear Studies,\\
High Energy Accelerator
Research Organization (KEK),\\
1-1, Oho, Tsukuba, Ibaraki 305-0801, Japan}
\address[label3]{School of Physics and Astronomy, Sun Yat-sen University,\\
Zhuhai, 519082, People's Republic of China}

%\begin{abstract}
%As is widely-known, the eigen-functions of the Landau problem in the
%symmetric gauge are specified by two quantum numbers. 
%The first is the familiar
%Landau quantum number $n$, whereas the second is the magnetic
%quantum number $m$, which is the eigen-value of the canonical
%orbital angular momentum (OAM) operator of the electron. 
%The eigen-energies of the system depend only on the first
%quantum number $n$, and the second quantum number
%$m$ does not correspond to any direct observables. 
%This appears to be reasonable,
%since the canonical OAM is in general a {\it gauge-variant}
%quantity, and observation of a gauge-variant quantity would
%contradict a fundamental principle of physics called the {\it gauge principle}.
%In recent papers, however, Bliohk et al. analyzed the motion of 
%electron helical beam along the direction of a uniform magnetic field,
%which was neglected in almost all the past analyses of the ordinary
%Landau states.
%According to them, while propagating along the direction of the
%magnetic field, the Landau electrons receive characteristic rotation
%with three different angular velocities, depending on the eigen-value
%$m$ of the canonical OAM operator, and this splitting was in fact experimentally
%confirmed. At first sight, this observation appears
%to contradict the gauge principle mentioned above.
%By making a through analysis of so-far only partially understood
%$m$-dependent rotational dynamics of the quantum Landau states, we try to 
%make clear how and why Bliohk et al.'s observation can be compatible with 
%the gauge principle. 
%\end{abstract}

\begin{abstract}
According to Bliokh et al., allowing free propagation along the direction of a 
uniform magnetic field, the familiar Landau electron state can be regarded 
as a non-diffracting version of the helical electron beam propagating along 
the magnetic field. Based on this observation, they argued that, while 
propagating along the magnetic field, the Landau electrons receive 
characteristic rotation with three different angular velocities, depending on
the eigen-value $m$ of the canonical OAM operator, which is generally 
gauge-variant, and this splitting was in fact experimentally confirmed.
Through complete analyses of highly mysterious
$m$-dependent rotational dynamics of the quantum Landau states, 
we try to make clear how and why their observation does not
contradict the widely-believed gauge principle. 
\end{abstract}

%%%%%%%%%%%%
% 117 words %
%%%%%%%%%%%%


\end{frontmatter}


%\section{}
%\label{}

\newpage
\section{Introduction}
\label{Section:s1}

The existence of propagating wave carrying intrinsic orbital angular
momentum (OAM) has been an object of intensive study and firmly 
established by now not only for photon 
beams but also for electron beams \cite{ABSW1992}\nocite{APB1999}
\nocite{TT2011}\nocite{BBSN2007}-\cite{Breview2017}. 
These helical (or twisted) beams are
characterized by an integer $m$ sometimes called the topological index of
the beam.
This integer is nothing but the eigen-value of the canonical
OAM operator, or more precisely its component along the propagating
direction of the photon or electron beam.
Although the canonical OAM is generally a gauge-variant quantity, its
observation does not contradict the famous gauge principle, just because
there is no difference between the canonical OAM and the manifestly
gauge-invariant mechanical (or kinetic) OAM for the free photon or electron
beam. However, this is not the
case for the recently-investigated helical electron beam propagating
under the influence of a uniform magnetic field \cite{BSVN2012},\cite{SSSLSBN}.
In the presence of non-zero magnetic field background, the two OAMs,
the gauge-variant canonical OAM and the gauge-invariant mechanical OAM
are absolutely different quantities, and they must be clearly distinguished.
Very interestingly, exactly the same problem also appears in a totally different 
field of physics. In fact, to clarify the difference between these two types
of OAMs inside the nucleon is one of the central issues of the so-called 
nucleon spin decomposition problem
in quantum chromo-dynamics \cite{LL2014},\cite{Waka2014}.

The purpose of the present paper is to carry out a complete analysis of
%so-far only partially understood 
very mysterious $m$-dependent rotational dynamics of
the Landau electron, by paying a special attention to highly
nontrivial role of the quantum guiding center in the Landau problem.
We also try to elucidate the difference between the two OAMs, i.e. the
gauge-variant canonical OAM and the gauge-invariant mechanical (or kinetic)
OAM in a nonzero electromagnetic field background. 
Our expectation is that these analyses would make clear how and why the 
quantum-number $m$-dependent splitting of the helical electron beam, 
while traveling along the direction of the uniform magnetic field, recently 
observed by Schattschneider et al. \cite{SSSLSBN}, can be compatible with the
widely-believed gauge principle as one of the fundamental principles of physics.

\section{Helical electron beam in a uniform magnetic field and Landau electron} 

Practically most important helical electron beam is the Laguerre-Gauss (LG) beam,
which is an approximate solution of free Helmholtz equation for
the electron in the paraxial approximation. 
Up to a normalization constant, the Laguerre-Gauss beam propagating
along the $z$-direction with the wave number $k$ is represented 
as \cite{ABSW1992}
\begin{eqnarray}
 \psi^{LG}_{n_r,m} (r, \phi, z) \ \propto \ 
 \left( \frac{r^2}{w^2 (z)} \right)^{|m| / 2} \,
 L^{|m|}_{n_r} \,\left( \frac{2 \,r^2}{w^2 (z)} \right) \, 
 e^{ \,\left( - \,\frac{r^2}{w^2 (z)} \,+ \,i \,k \,\frac{r^2}{2 \,R (z)} \right)} \,
 e^{\,i \,( m \,\phi \,+ \,k \,z)} \,e^{\,- \,i \,( 2 \,n_r \,+ \,|m| \,+ \,1 ) \,
 \arctan \,\left( \frac{z}{z_R} \right)} ,  
\end{eqnarray}
where $L^{|m|}_{n_r} (x)$ are the associated Laguerre polynomials,
$n_r = 0, 1, 2, \cdots$ is the number of radial nodes,
$w (z) = w_0 \,\sqrt{1 + z^2 \,/\,z_R^2}$ is the beam width depending on
$z$ due to diffraction, and $R (z) = z \,( 1 + z_R^2 \,/\, z^2)$ is
the radius of curvature of the wave front.
The transverse and the longitudinal scales of the beam are respectively
characterized by the waist $w_0$ (width in the focal plane $z = 0$)
and the Rayleigh difraction length $z_R$.
(Throughout the paper, we use the natural unit $\hbar = c = 1$.)

According to Bliokh et al. \cite{BSVN2012}, this LG beam is resembling 
the Landau states
of the electron in a $z$-directed uniform magnetic field $B$ in the
symmetric gauge represented as
\begin{equation}
 \psi_{n_r,m} (r, \phi, z) \ \propto \ 
 \left( \frac{r^2}{l_B^2} \right)^{|m| \,/\,2} \,L^{|m|}_{n_r} \,
 \left( \frac{r^2}{2 \,l_B^2} \right) \,e^{\,- \,\frac{r^2}{4 \,l_B^2}} \,\,
 e^{\,i \,( m \,\phi \,+ \,k_z \,z)} ,
\end{equation}
with the identification $w (z) \rightarrow 2 \,l_B$.
Here, $l_B \equiv 1 \,/\sqrt{e \,B}$ is the familiar magnetic length in 
the Landau problem. 
(In the present paper, the charge of the electron is taken to
be $- \,e$ with $e > 0$,
and the magnetic field $B \,(\, > 0)$ is assumed to be directed in 
the positive $z$-direction.)
As they argued, allowing free propagation along the magnetic field,
the Landau states represent non-diffracting versions of the 
electron helical beams. 

A remarkable observation by Bliokh et al. is that the rotation of electrons
in a uniform magnetic field in quantum picture is drastically different
from uniform classical orbiting, i.e. the familiar cyclotron motion.
Instead of rotation with a single cyclotron frequency 
$\omega_c = \frac{e \,B}{m_e}$, the Landau electrons, while propagating
along the direction of the magnetic field, receive characteristic rotation
with three different angular velocities, depending on the eigen-value $m$
of the canonical OAM operator $L_z^{can} = (\bm{r} \times \bm{p})_z$ : 
\begin{equation}
 \langle \omega \rangle \ = \ 
 \left\{ \begin{array}{ll}
 0 \ \ \ & \ \ \ (m < 0), \\
 \omega_L \ \ \ & \ \ \ (m = 0), \\
 \omega_c \ \ \ & \ \ \ (m > 0), \\
 \end{array} \right. \label{Eq:omega}
\end{equation}
where $\omega_c$ is the cyclotron frequency, while 
$\omega_L = \omega_c \,/\,2$ is the Larmor frequency.

We recall that above predictions are obtained by evaluating the
expectation value of the electron's angular velocity
$\omega (r) = v_\phi (r) \,/\,r$, with $v_\phi$ being the azimuthal
component of what-they-call the local Bohmian velocity given
by $\bm{v} = \bm{j} \,/\,|\psi|^2$. 
%(In the following, we assume that $\psi$ is normalized.)
Here, $\bm{j}$ is the familiar gauge-invariant probability current 
given by
\begin{equation}
 \bm{j} \ =  \ \frac{1}{m_e} \,\left[ \,\mbox{Im} \,
 ( \psi^* \,\nabla \,\psi ) \ + \ e \,\psi^* \bm{A} \psi \right] 
 \ = \ 
 \frac{1}{m_e} \,\mbox{Im} \,( \psi^* \bm{D} \,\psi ) ,
 \label{Eq:current}
\end{equation}
with $\bm{D} \equiv \nabla \,+ \,i \,e \,\bm{A}$ being the standard
covariant derivative.
Interestingly, the predicted $m$-dependent splitting of the
electron helical beam was later confirmed by a clever experiment
in which half of the beam is obstructed to stop with an opaque knife
edge stop and the spiral rotation of the visible part of the beam is
traced by moving the knife edge along the beam 
direction \cite{SSSLSBN}. 
This is really an interesting finding, and it motivated further
theoretical investigations in search of more complete understanding 
of the physics behind \cite{GSFB2014}\nocite{GFS2015}-\cite{SLSS2015}.

Despite those interesting researches, several questions remain.
First, the quantum number $m$ is the eigen-value of the
electron canonical OAM operator, which is usually believed
to be a gauge-variant quantity. Doesn't the observation of
$m$-dependent rotation contradict the well-known gauge principle,
which states that observables must be gauge-independent ?
Second, Bliokh et al. argue that the emergence of three different
types of rotation goes beyond simple classical picture of electron 
cyclotron motion in a uniform magnetic field, and it needs an
explanation based on quantum mechanics or the Bohmian 
mechanics \cite{DGTZ2004}.
Although the physical origin of the $m$-dependent splitting
of the electron's rotational motion was already discussed 
in their own perspectives \cite{BSVN2012},\cite{SSSLSBN}, here we 
can give a new and clearer insight into the problem based on 
the notion of {\it guiding center} known in the Landau problem.

\section{Landau electron's probability distributions and probability current
distributions}

To answer the questions raised in the previous section,
we point out that the following way of looking at the Landau problem is 
very useful. That is, we first recall the fact that, 
in the symmetric gauge $\bm{A} = \frac{1}{2} \,B \,( \,- \,y, x)$, the Landau
Hamiltonian $H = \frac{1}{2 \,m_e} \,(\bm{p} + e \,\bm{A})^2$ can
be expressed as a sum of the two pieces, i.e. the Hamiltonian of 
2-dimensional Harmonic oscillator and the
Zeeman terms \cite{DM1966}: 
\begin{equation}
 H \ = \ H_{osc} \ + \ H_{Zeeman} ,
\end{equation}
where
\begin{eqnarray}
 H_{osc} &=& \frac{1}{2 \,m_e} \,(p_x^2 + p_y^2) \ + \ 
 \frac{1}{2} \,m_e \,\omega_L^2 \, ( x^2 + y^2), \\
 H_{Zeeman} &=& \omega_L \,L_z^{can} .
\end{eqnarray}
Here, $\omega_L$ is the Larmor frequency, while $L_z^{can}$
is just the canonical OAM operator. 
The eigen-functions and the associated eigen-energies of 
the 2-dimensional Harmonic oscillator are well known.
They are given by
\begin{equation}
 H_{osc} \,\tilde{\psi}_{n_r,m} (r, \phi) \ = \ 
 (2 \,n_r + |m| + 1) \,\omega_L \,\,\tilde{\psi}_{n_r,m} (r,\phi),
\end{equation}
where
\begin{equation}
 \tilde{\psi}_{n_r,m} (r, \phi) \ = \ \frac{e^{\,i \,m \,\phi}}{\sqrt{2 \,\pi}} \,\,
 \tilde{R}_{n_r,m} (r) , \label{Eq:Landau_func1}
\end{equation}
with
\begin{equation}
 \tilde{R}_{n_r,n} (r) \ = \ \frac{1}{b} \,\sqrt{\frac{2 \,n_r !}{(n_r + |m|) !}} \,
 e^{\,- \,\frac{r^2}{2 \,b^2}} \,\left( \frac{r^2}{b^2} \right)^{|m| / 2} \,
 L^{|m|}_{n_r} \,\left( \frac{r^2}{b^2} \right) , \label{Eq:Landau_func2}
\end{equation}
and with $b^2 = 1 \,/\,(m_e \,\omega_L) = 2 \,/\,(e \,B)$.
In the above equations, $n_r \,(\,= 0, 1, 2, \cdots)$ represents the number of 
radial nodes, while $m$ does the magnetic quantum number,
which is the eigen-value of the canonical OAM operator
$L_z^{can} = - \,i \,\frac{\partial}{\partial \phi}$ : 
\begin{equation}
 L_z^{can} \, \tilde{\psi}_{n_r,m} (r, \phi) \ = \ 
 m \,\tilde{\psi}_{n_r,m} (r,\phi) ,
\end{equation}
with $m$ taking any integers.
Since $\tilde{\psi}_{n_r,m}$ are the simultaneous eigen-functions of $H_{osc}$
and $H_{Zeeman}$, it immediately follows that they are also the eigen-functions
of the whole Landau Hamiltonian,
\begin{equation}
 H \,\tilde{\psi}_{n_r,n} (r, \phi) \ = \ E \,\tilde{\psi}_{n_r,m} (r, \phi),
\end{equation}
with the corresponding eigen-energies,
\begin{equation}
 E \ = \ \left[\, (\,2 \,n_r + |m| + 1) + \ m \, \right] \,\omega_L.
\end{equation}

It is customary to introduce a new quantum number $n$ defined by
$n \equiv n_r + \frac{|m| + m}{2}$. This number takes zero or any positive
integer and it is called the Landau quantum number.
Accordingly, the eigen-functions of the Landau problem are standardly
expressed with $n$ and $m$ instead of $n_r$ and $m$, which motivates
to define new functions by
$\psi_{n,m} (r, \phi) \equiv \tilde{\psi}_{n_r,m} (r, \phi)$.
As a consequence, the eigen-energies of the Landau Hamiltonian depend
only on the quantum number $n$ as
\begin{equation}
 H \,\psi_{n,m} (r,\phi) \ = \ \left(\, 2 \,n + 1 \,\right) \,
 \omega_L \,\,\psi_{n,m} (r,\phi).
\end{equation}

These are all known stories, but the fact that the Landau eigen-states
are also the eigen-states of the 2-dimensional Harmonic oscillator
makes us notice an important symmetry of the eigen-functions.
First, remember that the radial wave functions $\tilde{R}_{n_r,m} (r)$ of
the 2-dimensional Harmonic oscillator have a simple symmetry
\begin{equation}
 \tilde{R}_{n_r, - \,m} (r)  \ = \ \tilde{R}_{n_r,m} (r)  ,
\end{equation}
i.e. the symmetry under the reverse of the magnetic quantum number $m$.
This symmetry comes from the time-reversal invariance of the
2-dimensional Harmonic oscillator Hamiltonian.
If this symmetry of $\tilde{R}_{n_r,m} (r)$ is translated into the symmetry of
the standard form of radial wave function in the Landau problem, defined as
$R_{n,m} (r) = \tilde{R}_{n_r,m} (r)$, we are led to a highly nontrivial
relation given by
\begin{equation}
 R_{n-m, \,- \,m} (r)  \ = \ R_{n,m} (r) .
\end{equation}

To understand surprising nature of this symmetry relation, let us,
for instance, consider the case where $n = m = 10$.
In this case, one has the relation $R_{0, - \,10} (r) = R_{10, 10} (r)$.
This means that the probability density of the state with $(n = 0, m = -\,10)$
is exactly the same as that of the state with $(n = 10, m = 10)$.
Note however that the eigen-energy of the former state
is $( 2 \times 0 + 1) \,\omega_L = \omega_L$, while that of the latter 
state is $( 2 \times 10 + 1) \,\omega_L = 21 \,\omega_L$.
We thus conclude that, though these two states have exactly the
same probability densities, they have totally different energies.
The resolution of this seeming paradox lies in the fact that, although
the probability densities of these two states are exactly the same,
they have totally different probability current distributions \cite{BSVN2012}.
One should recognize the fact that, under the presence of the external 
magnetic field, the internal electric current interacts with this magnetic field
so that this interaction also contributes to the energy of the system.

As seen from (\ref{Eq:current}), the gauge-invariant probability current 
consists of two pieces as
\begin{equation}
 \bm{j} \ = \ \bm{j}^{\,can} \ + \ \bm{j}^{\,gauge} ,
\end{equation}
with
\begin{equation}
 \bm{j}^{\,can} \ = \ \frac{1}{m_e} \,\mbox{Im} \,(\psi^* \,\nabla \,\psi ), \ \ \ \ 
 \bm{j}^{\,gauge} \ = \ \frac{1}{m_e} \,\psi^* \,e \,\bm{A} \,\psi ,
\end{equation}
which we hereafter call the canonical current and the gauge (potential)
current, respectively. 
(We recall that they are sometimes called the paramagnetic current and
the diamagnetic current \cite{GFS2015}.)
In the Landau states described by the eigen-functions (\ref{Eq:Landau_func1})
and (\ref{Eq:Landau_func2}), both have
only azimuthal components as $\bm{j}^{\,can} = j^{\,can}_\phi \,\bm{e}_\phi$
and $\bm{j}^{\,gauge} = j^{\,gauge}_\phi \,\bm{e}_\phi$, where
\begin{equation}
 j^{\,can}_\phi \ = \ \frac{1}{m_e} \,\,\frac{m}{r} \,\,\rho (r), \ \ \ \ \ 
 j^{\,gauge}_\phi \ = \ \frac{1}{m_e} \,\,\frac{r}{2 \,l_B^2} \,\,\rho (r) , 
 \label{Eq:currentphi}
\end{equation}
with $\rho (r) = |\psi|^2$ being the electron probability density.
Note that, due to the axial symmetry of the Landau eigen-states in
the symmetric gauge, $\rho$ is a function of $r$ only.

Also interesting is the angular momentum density $\bm{l}$
related to the probability current density $\bm{j}$ by
$\bm{l} = m_e \,\bm{r} \,\times \,\bm{j}$.
Note that this angular momentum $\bm{l}$ corresponds to the 
gauge-invariant mechanical (or kinetic) angular momentum $\bm{l}^{mech}$.
It has only $z$-component, and consists of two parts as
\begin{equation}
 l^{\,mech}_z \ = \ l^{\,can}_z \ + \ l^{\,gauge}_z , \label{Eq:l_can_gauge}
\end{equation}
where
\begin{equation}
 l^{\,can}_z \ = \ m \,\rho (r), \ \ \ \ \ 
 l^{\,gauge}_z \ = \ \frac{r^2}{2 \,l^2_B} \,\,\rho (r) .
\end{equation}

The canonical and the gauge parts of the OAM may also be called the
the paramagnetic OAM and the diamagnetic OAM, but 
we point out that the gauge part of the OAM is nothing 
but what-we-called the potential angular momentum $l^{pot}_z$ in the 
paper \cite{Waka2010} aside from the
sign difference, i.e. $l^{\,gauge} = - \,l^{\,pot}_z$.
(There is a reason in this sign convention in the definition of the
potential angular momentum. The potential angular momentum
is contained in the expression of the total photon angular momentum
given by $\int \,\bm{r} \times (\bm{E} \times \bm{B}) \,d^3 x$
in the interacting system of photons and charged particles, so that
it has a meaning of the angular momentum carried by the electromagnetic
field in the presence of the charge particles.)
We prefer to use the terminology potential OAM instead of diamagnetic
OAM, because it has a universal meaning in the general theory of 
electromagnetism as well as in quantum chromodynamics as a nonabelian 
gauge theory. (See \cite{Waka2010} or \cite{WKZZ2018} for more details.)

We recall that spatial integrals of these quantities, which are just
the expectation values of the corresponding operator in the Landau
state $\psi_{n,m}$, are well-known. They are given by 
\cite{BSVN2012},\cite{WKZ2018},\cite{FL2000}
\begin{equation}
 \langle \,l^{\,can}_z \,\rangle \ = \ m, \ \ \ \ \ 
 \langle \,l^{\,gauge}_z \,\rangle \ = \ - \,
 \langle \,l^{\,pot}_z \,\rangle \ = \ 
 2 \,n + 1 - m ,
\end{equation}
so that we have
\begin{equation}
 \langle \,l^{\,mech}_z \,\rangle \ = \ \langle \,l^{\,can}_z \,\rangle \ - \ 
 \langle \,l^{\,pot}_z \,\rangle \ = \ 2 \,n \ + \ 1.
\end{equation}
This means that the expectation value of the mechanical
OAM operator depends only on the Landau quantum number $n$.

%%%%%%%%%%%%%%%%%%%%%%%%%%%%%%%%%%%%%%%%%%%%%%%%%%%%%%%%%%%%
\begin{figure*}[ht]
\begin{center}
\includegraphics[width=12.5cm]{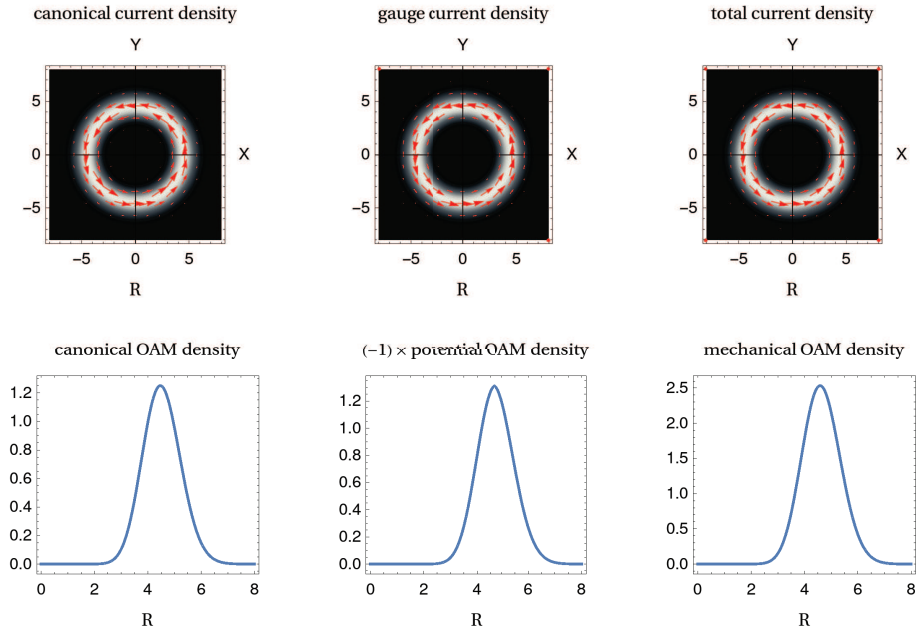}
\caption{The three figures in the upper panel respectively represent the  
distributions of the canonical current, the gauge current, and the total current 
(red arrows in color) together with the probability distribution of
the electron corresponding to the Landau eigen-state specified by the
quantum numbers $n_r = 0$ and $m = + \,10$. The three figures in the lower 
panel represent the corresponding canonical OAM density, the potential OAM 
density $\times \,(-1)$,
% (aside from the extra negative sign), 
and the total OAM densities,
respectively. Here, the dimensionless coordinates $X = x/l_B$, $Y = y / l_B$
and $R = r / l_B$ are used.} 
\label{Fig:fig1}
\end{center}
\end{figure*}
%%%%%%%%%%%%%%%%%%%%%%%%%%%%%%%%%%%%%%%%%%%%%%%%%%%%%%%%%%%%

Just for completeness, we point out that the electron's angular velocity
operator $\omega (r) = j_\phi (r) / r$ is also given as a sum of the
contributions of the canonical current and of the gauge current as
\begin{equation}
 \omega (r) \ = \ \frac{1}{m_e} \,\frac{m}{r^2} \,\rho (r) \ + \ 
 \frac{1}{m_e} \,\frac{1}{2 \,l_B^2} \,\rho (r) .
\end{equation}
Evaluating its expectation value in the Landau state with use of
the relation $\langle \rho (r) / r^2 \rangle = 
1 / \left(2 \,l_B^2 \,|m| \right)$
as well as $\langle \rho (r) \rangle = 1$, we get
\begin{equation}
 \langle \omega (r) \rangle \ = \ \omega_L \,
 \left( \frac{m}{|m|} \ + \ 1 \right) , \label{Eq:omegaeq}
\end{equation}
which confirms the relation (\ref{Eq:omega}). 
We point out that this relation was
already written down in the paper by Li and Wang \cite{LW1999}, although its
practical importance became clear only after the proposal of
using the helical electron beams \cite{BSVN2012},\cite{SSSLSBN}.

%%%%%%%%%%%%%%%%%%%%%%%%%%%%%%%%%%%%%%%%%%%%%%%%%%%%%%%%%%%%
\begin{figure*}[ht]
\begin{center}
\includegraphics[width=12.5cm]{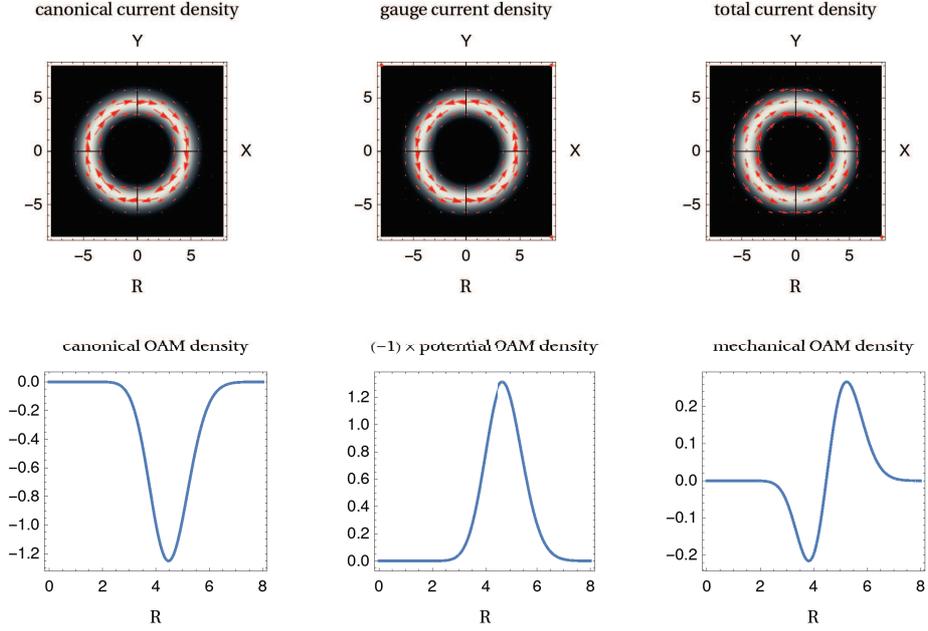}
\caption{The same as the Fig.\ref{Fig:fig1} but for the Landau eigen-states
specified by the quantum numbers $n_r = 0$ and $m = - \,10$.}
\label{Fig:fig2}
\end{center}
\end{figure*}
%%%%%%%%%%%%%%%%%%%%%%%%%%%%%%%%%%%%%%%%%%%%%%%%%%%%%%%%%%%%%

In Fig.1 and Fig.2, we show the probability current densities together with 
the probability densities, and also the angular momentum densities.
Fig.1 corresponds to the state with $(n_r = 0, m = 10)$ or equivalently
$(n = 10, m = 10)$, while Fig.2 to the state with $(n_r = 0, m = - \,10)$
or $(n=0, m = - \,10)$.
In the three figures on the upper panel, the higher probability density region
is drawn by brighter (white) color, whereas the lower density region is
by darker (black) color.
One can confirm that the probability density of the state with
$(n = 10, m = 10)$ shown in Fig.1 and that with $(n = 0, m = - \,10)$
shown in Fig.2 are exactly the same, in spite that their eigen-energies are
totally different. However, the probability current densities shown
by arrows (red in color) are entirely different for these two states.
Since $m > 0$ for the state with $(n = 10, m = 10)$, both of $j^{can}_\phi$
and $j^{gauge}_\phi = - \,j^{pot}_\phi$ are positive, which means that 
canonical current as well as the gauge current are circulating in a 
counter-clock-wise direction. 
Accordingly, the total current is also flowing counter-clock-wise. 
(See the three figures on the left panel of Fig.1.)
On the other hand, since $m < 0$ for the state with $(n = 0, m = - \,10)$, the
canonical current is flowing counter-clock-wise, whereas the gauge current
is circulating clock-wise. Because of different radial dependencies of the
canonical and gauge currents i.e. $j^{can}_\phi (r) \propto \frac{1}{r} \,\rho (r)$
and $j^{gauge}_\phi (r) \propto r \,\rho (r)$, the flow of the total current
shows highly nontrivial behavior as illustrated in the third figure
on the upper panel of Fig.2 \cite{BSVN2012}.
That is, the flow of the net current is counter-clock-wise in the outer
part of the high probability density region, whereas it is clock-wise
in the inner part of high probability density region.

On the lower panel of Fig.1 and Fig.2, we show the radial dependencies
of the canonical OAM density, $(- \,1) \,\times$ potential OAM density,
and the mechanical OAM density. The behaviors of the OAM densities
illustrated on the lower panel of these figures are easily understood
from those of the corresponding probability current densities illustrated on
the upper panel of Fig.1 and Fig.2.
Note that, when integrated over the whole space, the mechanical 
OAM always takes the value $(2 \,n + 1)$ irrespectively of the
value of the magnetic quantum number $m$.
These features, which are automatic consequences of the eigen-solutions
of the Landau Hamiltonian, were already pointed out in the paper 
by Bliokh et al. \cite{BSVN2012}.
However, our explanation based on the notion of guiding center
gives much clearer physical explanation about why the probability 
current distributions show such nontrivial behaviors in dependence of the 
value of $m$.
Moreover, as we shall see below, our explanation above also help us 
to get clearer 
understanding of the novel splitting phenomena of the electron 
helical beam into three pieces depending on the eigen-value $m$ of 
the canonical OAM operator.

\section{Explanation of $m$-dependent splitting of
helical electron beam in a magnetic field}

What plays an important role in giving a clear physical interpretation
on the strange $m$-dependent splitting
of the helical electron beam in a uniform magnetic field is the notion 
of guiding center introduced by Johnson and Lippmann
many years ago \cite{JL1949}.
For readers who are not familiar with the concept of guiding center, 
we think it helpful to recall its basic properties.
In classical mechanics, the motion of a electron with charge 
$- \,e \,(e > 0)$ and the mass $m_e$ is determined by the classical 
equation of motion
\begin{equation}
 m_e \,\dot{\bm{v}} (t) \ = \ - \,e \,\left( \bm{v} (t) \times \bm{B} \right),
\end{equation}
where the dot stands for the time derivative and $\bm{v} (t) = \dot{\bm{x}} (t)$
is the electron's velocity.
The solution for the electron's orbit $(x (t), y (t))$ with the initial conditions
$x (0) = x_0, y (0) = y_0, v_x (0) = v_{x0}, v_y (0) = v_{y0}$ is easily obtained as
\begin{eqnarray}
 x (t) &=& X \ + \ \frac{1}{\omega_c} \,v_y (t), \hspace{8mm}
 X \ = \ x_0 \ - \ \frac{v_{x0}}{\omega_c}, \\
 y (t) &=& Y \ - \ \frac{1}{\omega_c} \,v_x (t), \hspace{8mm}
 Y \ = \ y_0 \ - \ \frac{v_{y0}}{\omega_c},
\end{eqnarray}
where $v_x (t) = v_0 \,\cos (\omega_c \,t + \alpha), 
v_y (t) = v_0 \,\sin (\omega_c \,t + \alpha), v_0 = \sqrt{v^2_{x0} + v_{y0}^2}$,
and $\tan \alpha = v_{y0}\,/\,v_{x0}$.
Here, the quantity $(X, Y)$ has a clear physical meaning as the center of
cyclotron motion. Obviously, the center coordinates $(X, Y)$ of the cyclotron
motion is time independent, $\dot{X} = \dot{Y} = 0$.
We also realize from the above solution that the following two quantities
are constants in time : 
\begin{eqnarray}
 r^2_c &\equiv& (x (t) - X)^2 \ + \ (y (t) - Y)^2 \ = \ 
 \frac{m_e^2 \,v_0^2}{e^2 \,B^2}, \\
 R^2 &\equiv& X^2 + Y^2,
\end{eqnarray}
where $r_c$ represents the cyclotron radius, while $R$ does the distance 
between the coordinate origin and the center of cyclotron motion.

When going to quantum theory, the centroid $(X, Y)$ of the cyclotron motion
is called the guiding center and its physical meaning becomes less intuitive as
compared with the classical case.
In fact, in quantum mechanics, the mechanical momentum $m_e \,\bm{v}$
is replaced by an operator $\hat{\bm{\Pi}} = - \,i \,\nabla + e \,\bm{A}$,
and consequently the guiding center coordinates also becomes quantum 
operators as
\begin{eqnarray}
 \hat{X} &=& x \ - \ \frac{1}{e \,B} \,\hat{\Pi}_y \ = \ 
 x \ - \ \frac{1}{e \,B} \,
 \left[ - \,i \,\frac{\partial}{\partial y} \ + \ e \,A_y \right], \\
 \hat{Y} &=& x \ + \ \frac{1}{e \,B} \,\hat{\Pi}_x \ = \ 
 y \ + \ \frac{1}{e \,B} \,
 \left[ - \,i \,\frac{\partial}{\partial x} \ + \ e \,A_x \right].
\end{eqnarray}
Here, we add hat symbols to $(X, Y)$ and $(\Pi_x, \Pi_y)$ to emphasize
that they are quantum operators, although we shall omit them below for
notational simplicity. Note that, even in quantum mechanics, the guiding 
center coordinates $X$ and $Y$ are still constants of motion, since they
commute with the Landau Hamiltonian $H$, i.e. $[X, H] = [Y, H] = 0$.
It also holds that $[R^2, H] = 0$ with $R^2 = X^2 + Y^2$.
However, the two $q$-numbers $X$ and $Y$ do not commute with
each other. They rather satisfy the commutation relation
$[X, Y] = i \,l_B^2$ with $l_B$ the magnetic length.
This means that we cannot specify the $x$- and $y$- coordinates
of the guiding center simultaneously with arbitrary precision.
(We point out that quantum-mechanically nontrivial role of
the guiding center in the Landau problem was also discussed 
in the two recent papers \cite{KWZZ2019},\cite{Enk2019}
from a different perspective.)
We shall later see that this quantum mechanical uncertainty in the
position of the guiding center coordinates plays a decisively important 
role for understanding highly nontrivial structure of the probability current
distribution of the electron in the Landau problem.

In quantum mechanics, the cyclotron radius $r_c$ also becomes a
quantum operator, which is sometimes called the orbit radius operator.
As pointed out by Johnson and Lippmann many years 
ago \cite{JL1949}, $r^2_c$ is related to the Landau Hamiltonian or 
the system energy as
\begin{equation}
 H \ = \ \frac{1}{2} \,m_e \,( v_x^2 \ + \ v_y^2 ) \ = \ 
 \frac{1}{2} \,m_e \,\left( \frac{e \,B}{m_e} \right)^2 \,\,
 \left\{ (x - X)^2 \ + \ (y - Y)^2 \right\} \ = \ 
 \frac{1}{2} \,m_e \,\omega^2_c \,r^2_c ,
\end{equation}
so that it is a constant of motion also in quantum mechanics.

Johnson and Lippmann also pointed out that, $R^2$ and $r^2_c$
satisfy the following nontrivial relation : 
\begin{equation}
 L^{can}_z \ = \ \frac{1}{2 \,l_B^2} \,( r^2_c \ - \ R^2 ) , 
 \label{Eq:L_can_rc_R}
\end{equation}
where $L^{can}_z$ is the canonical OAM operator.
The expectation values of the above quantities in the Landau
eigen-state $\psi_{n,m}$ (or $\tilde{\psi}_{n_r, m}$) can easily be
evaluated as \cite{FL2000},\cite{LW1999}
\begin{eqnarray}
 \langle r^2_c \rangle &=& 2 \,
 \left( n_r \ + \ \frac{|m| + m}{2} \ + \ \frac{1}{2} \right) \,l^2_B
 \ = \ (2 \, n \ + \ 1) \,l^2_B, \\
 \langle R^2 \rangle &=& 2 \,
 \left( n_r \ + \ \frac{|m| - m}{2} \ + \ \frac{1}{2} \right) \,l^2_B
 \ = \ (2 \, n \ - \ 2 \,m \ + \ 1) \,l^2_B ,
\end{eqnarray}
which gives
\begin{equation}
 \langle L^{can}_z \rangle \ = \ m , \label{Eq:L_can_m}
\end{equation}
as naturally anticipated. From Eqs.(\ref{Eq:L_can_rc_R}) and 
(\ref{Eq:L_can_m}), the following relation immediately follows : 
\begin{equation}
 \left\{ \begin{array}{lll}
 \ \sqrt{\langle r_c^2 \rangle} \ \ > \ 
 \sqrt{\langle R^2 \rangle} \ \ & \ \mbox{when} \ & \ m > 0, \\
 \ \sqrt{\langle r_c^2 \rangle} \ \ = \ 
 \sqrt{\langle R^2 \rangle} \ \ & \ \mbox{when} \ & \ m = 0, \\
 \ \sqrt{\langle r_c^2 \rangle} \ \ < \ 
 \sqrt{\langle R^2 \rangle} \ \ & \ \mbox{when} \ & \ m < 0 .
 \end{array} \right.
\end{equation}
Thus, one realizes that the sign of the magnetic quantum number $m$
is inseparably connected with the magnitude correlation between
$r_c$ and $R$.

It is instructive to compare once again the two typical states, i.e. the
state with $(n, m ) = (10, 10)$ and that with $(n, m) = (0, - \,10)$.
For the former state, we have $\sqrt{\langle r_c^2 \rangle} = 
\sqrt{21} \,l_B$ and $\sqrt{\langle R^2 \rangle} = l_B$,
while for the latter state, we have $\sqrt{\langle r_c^2 \rangle} = 
l_B$ and $\sqrt{\langle R^2 \rangle} = \sqrt{21} \,l_B$.
Thus, for the state with $(n, m) = (10, 10)$, the Landau electron
is making a circular motion with the radius of $\sqrt{21} \,l_B$
around the guiding center which lies inside the circle of radius
$l_B$, as schematically illustrated on the left panel of Fig.3.
On the other hand, for the state with $(n, m) = (0, - \,10)$,
the electron is rotating with the radius of $l_B$ around the
guiding center which is located on the circle of radius $\sqrt{21} \,l_B$
as illustrated on the right panel of Fig.3.
Note that, in quantum mechanics, the position of the guiding
center is inherently uncertain and it is distributed on the circle of
radius $\sqrt{21} \,l_B$ with equal probability.
For this reason, the quantum mechanical probability
distribution $\rho$ of the electron as well as its probability
current distribution $\bm{j}$ are destined to have {\it axial symmetries}
around the coordinate origin in consistent with their forms
already shown in Fig.1 and Fig.2. In particular, from the right
panel of Fig.3, one can clearly understand the reason why the
flow of the net current for the state with negative $m$ is
counter-clock-wise in the outer part of the high probability
density region, while it is clock-wise in the inner part
of high probability density region. This transparent explanation
on the characteristic structure of the probability current distribution for
the Landau electron on the basis of the concept of the quantum
guiding center is one of our main findings.

%%%%%%%%%%%%%%%%%%%%%%%%%%%%%%%%%%%%%%%%%%%%%%%%%%%%%%%%%%%%%
\begin{figure*}[ht]
\begin{center}
\includegraphics[width=9cm]{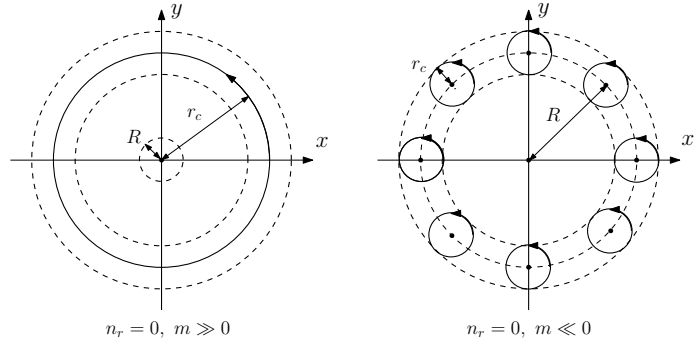}
\caption{Schematic pictures of the quantum mechanical cyclotron motion
of the electron in a uniform magnetic field. The left figure corresponds to
the case where the node number $n_r$ of the radial wave function is zero, while
the magnetic quantum number $m$ is large and positive.
On the other hand, the right figure corresponds to the case where $n_r = 0$,
while $m$ is largely negative. In both figures, $r_c$ represents the radius of
the cyclotron motion, whereas $R$ does the distance between the guiding
center and the coordinate origin. Note that the position
of the guiding center is statistically distributed on the circle of radius $R$
with equal probability.}
\label{Fig:fig3}
\end{center}
\end{figure*}
%%%%%%%%%%%%%%%%%%%%%%%%%%%%%%%%%%%%%%%%%%%%%%%%%%%%%%%%%%%%%%%%%

The comparison of the two panels in Fig.3 also provides us
with a clear explanation on the $m$-dependent splitting of
electron's rotational trajectory while propagating along the
direction of the magnetic field. For the state with $m > 0$,
the electron is certainly rotating around the origin with the
cyclotron frequency $\omega_c$. On the other hand, for
the state with $m < 0$, the electron is not actually rotating
around the origin, which explains the reason why its angular velocity
$\langle \omega \rangle$ equals to zero. 
 (See also the paper by Li and Wang \cite{LW1999}.)

%A slightly sophisticated is
%the $m = 0$ case. Bliokh et al. emphasized that the rotational
%frequency corresponding to this state just coincides with the Larmor
%frequency and suggested as if the appearance of the Larmor frequency
%must have some deep physical meaning \cite{BSVN2012},\cite{SSSLSBN}.
%In our opinion, however, there is no surprise in the appearance of the 
%Larmor frequency here. In fact, the $m = 0$ state can conveniently be
%expressed as a superposition of the states $|\,n, m_0 \rangle$
%and $|\,n, - \,m_0 \rangle$, i.e. as
%$|\,n, m = 0 \rangle = \frac{1}{\sqrt{2}} \,\left\{ |\,n, m = m_0 \rangle
%+ |\,n, m = - \,m_0 \rangle \right\}$, with any Landau quantum number $n$
%and with any positive integer $m_0$. 
%(In fact, it is an easy exercise to verify that the expectation value of
%$L^{can}_z$ in this state is zero.) Evaluating the expectation value of
%the angular velocity operator $\omega (r)$ in this state, we get
%
%\begin{equation}
% \langle \omega (r) \rangle \ = \ \frac{1}{2} \,\left\{\, 
% \langle \omega (r) \rangle_{m = m_0} \ + \ 
% \langle \omega (r) \rangle_{m = - \,m_0} \,\right\}
% \ = \ \frac{1}{2} \,\left( \omega_c \ +  0 \right) \ = \ \omega_L .
%\end{equation}
%
%Thus we see that the Larmor frequency in the $m = 0$ case
%arises, simply because it is the average of the cyclotron frequency
%$\omega_c$ in the $m > 0$ case and the zero frequency in the
%$m < 0$ case.

A slightly delicate is the $m=0$ case. It corresponds to the 
situation $\sqrt{\langle r_c^2 \rangle} = \sqrt{\langle R^2 \rangle}$,
which means that the most probable trajectory of the electron's
cyclotron motion passes through the coordinate origin.
Bliokh et al. emphasized that the angular velocity
corresponding to this mode coincides with the Larmor frequency 
$\omega_L$ and suggested as if the appearance of the Larmor frequency 
has some deep reason \cite{BSVN2012},\cite{SSSLSBN}.
In our opinion, there is no mystery in the appearance of the Larmor
frequency here. To understand it, it is simpler to go
back to the formula (\ref{Eq:omegaeq}). 
The 1st and the 2nd terms on the right-hand side of
this equation represent the contributions of the canonical
current and the gauge current to the angular velocity
$\langle \omega (r) \rangle$. The gauge current contribution equals 
to the Larmor frequency $\omega_L$ irrespectively of the value
of $m$. On the other hand, the canonical current contribution is
$\pm \,\omega_L$ depending on the sign of $m$.
Then, for the $m > 0$ mode, these two contributions
are added up to give $2 \,\omega_L = \omega_c$, i.e. the
cyclotron frequency. On the other hand, for the $m < 0$ mode,
these two contributions are exactly canceled out to give zero
rotational velocity in conformity with the schematic picture illustrated
on the right panel of Fig.3.
Finally, for the marginal case of $m=0$, the gauge current contribution
is still $\omega_L$, but the canonical current contribution vanishes, as is
clear from the expression (\ref{Eq:currentphi}) for the canonical current.
Then, it can alternatively be said that the Larmor frequency for the
$m=0$ mode appears just because it is an average of the two frequencies 
$\omega_c$ and $0$ corresponding to the two types of cyclotron motions, 
i.e. the one which rotates around the origin with the frequency $\omega_c$
and the other which does not actually rotate around the origin.

\section{Conclusion}

To sum up, we have carried out a comprehensive analysis of the 
$m$-dependent rotational dynamics of the Landau eigen-states
$|\,n, m \rangle$ in the symmetric gauge and confirmed that 
unexpectedly rich structure is hidden in its $m$-dependencies.
They are the novel symmetry of the electron's
probability densities of the two Landau states $|\,n-m, - \,m \rangle$
and $|\,n, m \rangle$ and also the highly nontrivial structure of the
probability current distribution, which critically depends on the sign of the
quantum number $m$. 
In particular, we demonstrated that the above-mentioned  nontrivial 
structure of the probability current distribution has a simple
intuitive explanation based on the unique role of the quantum
guiding center concept in the Landau problem.
The novel $m$-dependent splitting of the electron's rotational
motion, while propagating along the direction of the magnetic
field, can also be transparently understood if we notice the magnitude 
correlation between the cyclotron radius and the distance of
the guiding center from the coordinate origin, which critically depends
on the sign of $m$.
Since this $m$-dependent splitting of the electron's rotational
trajectory is a prediction based on the gauge-invariant total or
mechanical current, it never contradicts the gauge principle.
Rather, the remaining degeneracy of the rotational
frequency $\langle \omega \rangle$ for both of the $m>0$ mode 
and of the $m<0$ mode may be interpreted as a consequence
of the gauge-invariance requirement for observables.

\vspace{1mm}
\noindent
%{\bf Acknowledgement}
\section*{Acknowledgment}

\vspace{1mm}
\noindent
M.~W. thanks the Institute of Modern Physics of the Chinese
Academy of Sciences in Lanzhou for hospitality.
Y.~K. L-P.~.Z and P.-M.~Z. are supported by the National Natural
Science Foundation of China (Grant No.11575254 and No.11805242).
This work is partly supported by the Chinese Academy of Sciences
President's International Fellowship Initiative 
(No. 2018VMA0030 and No. 2018PM0028).

%\bibliography{mybibfile}



%\vspace{3mm}
%\appendix

\vspace{1mm}
\noindent
\section*{References}

\bibliographystyle{elsarticle-num}
%% Numbered
\bibliography{mybibfile}


\end{document}